\documentclass[final,5p,times,twocolumn]{elsarticle}

\usepackage[colorlinks=true,linkcolor=black,citecolor=blue,urlcolor=blue,bookmarks=false]{hyperref}
\hypersetup{breaklinks=true}
\usepackage[utf8]{inputenc}
\usepackage{bm,mathtools,cleveref}
\usepackage{array}
\usepackage{epsfig}
\usepackage{amsmath}
\usepackage{amssymb}
\usepackage{multirow}
\usepackage{slashed}
\usepackage{subcaption}
\usepackage{comment}
\usepackage{enumitem}
\graphicspath{{./fig/}}

\usepackage[normalem]{ulem}
\usepackage[x11names]{xcolor}

\renewcommand{\sout}{\bgroup \color{red} \ULdepth=-.5ex \ULset}

\journal{Physics Letters B}
\biboptions{sort&compress}

\begin{document}

\begin{frontmatter}

\title{One-nucleon intermediate-state contribution to the pion-coupled four-quark condensate in the $N$--$J/\psi$ interaction}

\author[yonsei]{Seokwoo Yeo}
\ead{ysw351117@gmail.com}

\author[jaea]{Philipp Gubler}
\ead{philipp.gubler1@gmail.com}

\author[yonsei]{Su Houng Lee}
\ead{suhoung@yonsei.ac.kr}

\affiliation[yonsei]{
organization={Department of Physics and Institute of Physics and Applied Physics, Yonsei University},
city={Seoul},
postcode={03722},
country={Korea}
}

\affiliation[jaea]{
organization={Advanced Science Research Center, Japan Atomic Energy Agency},
city={Tokai, Naka, Ibaraki},
postcode={319-1195},
country={Japan}
}

\begin{abstract}
The density dependence of four-quark condensates is a major source of uncertainty in operator product expansion analyses of hadrons in nuclear matter. Their determination is particularly relevant to the $N$--$J/\psi$ interaction, whose long-range attraction appears to be well described by two-pion exchange in lattice QCD. We evaluate the pion-coupled four-quark condensates using one-nucleon intermediate-state insertions and incorporate them into the QCD sum rule analysis for the $J/\psi$ meson in nuclear matter. At normal nuclear matter density, we find a mass shift of $-1.75$ MeV, which is of the same order as the approximately $-5$ MeV attraction inferred from a two-pion exchange potential fitted to lattice QCD results. A previous factorization-based treatment required the introduction of unnaturally large phenomenological parameters to obtain a comparable shift. Our result shows that one-nucleon intermediate states provide a significant contribution to the in-medium four-quark condensates and improve the connection between the operator product expansion and the long-range $N$--$J/\psi$ interaction.
\end{abstract}

\end{frontmatter}

\section{Introduction}
The $N$--$J/\psi$ interaction provides a useful probe of nonperturbative QCD. Because the $J/\psi$ contains no valence light quarks, its coupling to a nucleon is expected to arise mainly from gluonic
interactions~\cite{Peskin1979,BrodskyEtAl1990}. At long distances, however, such gluonic interactions manifest themselves through color-singlet hadronic degrees of freedom.

Developments in lattice QCD have made it possible to extract hadron--hadron interaction potentials~\cite{AokiEtAl2012}. While physical observables can be obtained from these potentials, their spatial behavior can also provide insight into the underlying
interaction mechanisms. In fact, a lattice QCD study of the $N$--$J/\psi$ system finds an
attractive interaction with a long-range tail consistent with two-pion
exchange~\cite{LyuEtAl2025}. This indicates that light-quark and chiral dynamics also contribute to the interaction. Such behavior was anticipated in a color-dipole analysis, where correlated two-Goldstone-boson exchange gives the leading long-range contribution~\cite{FujiiKharzeev1999}. A similar two-pion tail was found in the lattice $N$--$\phi$ potential~\cite{LyuEtAl2022}, supporting a broader two-pion exchange
picture for long-range nucleon--hadron interactions in which one-pion exchange is forbidden~\cite{Hatsuda2026}. A dispersive analysis likewise shows that low-energy $N$--$J/\psi$ scattering is dominated by soft-gluon exchange through correlated light-meson channels~\cite{WuEtAl2025}.

QCD sum rules relate hadron properties to QCD condensates~\cite{SVZ1979,ReindersEtAl1985}. The generalization of the sum rule approach to vector mesons in nuclear matter was discussed in Ref.~\cite{HatsudaLee1992}. For the $J/\psi$ in nuclear matter, early
analyses included density-dependent dimension-four scalar and twist-2
gluon condensates~\cite{KlinglEtAl1999,Hayashigaki1999}, and later work
considered dimension-six nonzero-spin gluon operators~\cite{KimLee2001}.
The effect of in-medium $D$-meson mass modifications on charmonium masses was also estimated through $D\bar D$ loop contributions~\cite{LeeKo2003}.

Recently, two of us investigated whether the long-range two-pion exchange contribution could also be obtained within the operator product expansion (OPE) framework~\cite{YeoEtAl2025}. This was done by extracting the leading four-quark operator with the structure of two pseudoscalar or axial-vector currents, which is expected to couple directly to two pions. The nucleon expectation value, however, was obtained using the ground-state saturation approximation (GSA), as discussed below. Such an approximation is problematic unless the four-quark operator is a chiral order parameter. Indeed, the resulting mass shift was attractive, in qualitative agreement with lattice QCD~\cite{LyuEtAl2025,YeoEtAl2025}, but its magnitude was found to be smaller than that inferred from the lattice-fitted two-pion exchange potential. A comparable shift required relatively large values of $\kappa_{\Delta\rho}$, which parameterizes deviations from the GSA, and of the infrared strong coupling. The QCD sum rule result, therefore, did not provide a completely satisfactory description of the $N$--$J/\psi$ interaction.

The largest quantitative uncertainty in QCD sum rules stems from the treatment of four-quark condensates. Their matrix elements are commonly estimated using the GSA, also known as factorization, which approximates four-quark condensates by products of ground-state two-quark condensates~\cite{Leupold2005}. Studies of chiral-partner sum rules have shown, however, that chirally symmetric four-quark components not captured by the GSA can be nearly equal in magnitude to their chiral-symmetry-breaking counterparts~\cite{KimLee2021,KimLee2022}. Furthermore, the limitations of factorization may become more pronounced in nuclear matter, where physical intermediate states contribute to nucleon matrix elements of four-quark operators.

In this work, we improve the estimate of the four-quark condensates in symmetric nuclear matter within the linear-density approximation by inserting a complete set of physical states between the two quark bilinears and retaining the one-nucleon contribution. Baryon-number conservation restricts the intermediate states to those with baryon number one, and the nucleon is the lowest such state. Contributions associated with higher baryonic excitations are expected to be suppressed by transition form factors, while the low-lying $N\pi$
continuum may provide an additional contribution to the pion-coupled operators. We decompose the relevant operators into their isoscalar and isovector components and evaluate the corresponding nucleon matrix elements using momentum-dependent form factors. We find that the one-nucleon estimate of the four-quark condensate has the same overall sign as the ground-state saturation estimate but is larger in magnitude. We then incorporate these matrix elements into the QCD sum rule for the $J/\psi$, using the OPE inputs and sum rule setup of Ref.~\cite{YeoEtAl2025}, unless stated otherwise. This treatment brings the predicted mass shift closer to that inferred from the lattice-fitted two-pion exchange potential without requiring the unnaturally large parameter values used in the previous analysis.

This paper is organized as follows. In Section \ref{fourquark}, we introduce and develop our framework accounting for the contribution of one-nucleon intermediate states to the nuclear expectation values of four-quark condensates, using nucleon form factors from lattice QCD. In Section \ref{qsr}, we describe the OPE and QCD sum rule analysis of the $J/\psi$ in nuclear matter, and in Section \ref{qsr_results}, we present the numerical results. The paper is concluded with a summary in Section \ref{summary}.

\section{Four-quark condensates from one-nucleon intermediate states}
\label{fourquark}

\subsection{Pion-coupled four-quark operators}
\label{subsec:pion_operators}

From the OPE point of view, the two-pion exchange contribution is expected to come from the four-quark operator that has the structure of two pseudoscalar or axial-vector currents. The nucleon expectation values of these operators then determine the corresponding OPE contribution to the $N$--$J/\psi$ interaction. In Ref.~\cite{YeoEtAl2025}, we derived the leading-order contribution to these operators from the dimension-six gluon operators.

The dimension-six gluon operators entering the $J/\psi$ OPE can be related to four-quark operators by means of the gluon equation of motion,
\begin{align}
D^\nu G^a_{\mu\nu}
&=
\frac{g_s}{2}
\sum_{f=u,d,s}
\bar q_f\gamma_\mu\lambda^a q_f .
\end{align}
In particular, the scalar operator can be written as
\begin{align}
\left(D^\alpha G^a_{\mu\alpha}\right)
\left(D_\beta G^{a\mu\beta}\right)
&=
\frac{g_s^2}{4}\mathcal O_{4q},
\
\mathcal O_{4q}
\equiv
\sum_{f,g}
\left(\bar q_f\gamma_\mu\lambda^a q_f\right)
\left(\bar q_g\gamma^\mu\lambda^a q_g\right).
\end{align}

We first review the evaluation of the four-quark condensate in vacuum. To make the color, Dirac, and flavor indices explicit, we write
\begin{align}
\mathcal O_{4q}
={}&
\sum_{f,g}
\bar q_{f,r}^{\alpha}
(\gamma_\mu)_{rs}
(\lambda^a)_{\alpha\beta}
q_{f,s}^{\beta}
\,
\bar q_{g,t}^{\gamma}
(\gamma^\mu)_{tu}
(\lambda^a)_{\gamma\delta}
q_{g,u}^{\delta},
\end{align}
where $\alpha,\beta,\gamma,\delta$ denote color indices in the fundamental representation and
$r,s,t,u$ Dirac indices.
We assume that each quark field is paired with an antiquark field, with the corresponding quark bilinears saturated by the vacuum. The presence of two quark and two antiquark fields then allows two possible contractions, leading to 
\begin{align}
\langle \mathcal O_{4q}\rangle_0
\simeq
\sum_{f,g}
&(\gamma_\mu)_{rs}
(\lambda^a)_{\alpha\beta}
(\gamma^\mu)_{tu}
(\lambda^a)_{\gamma\delta}
\bigg[
\langle 0|
\bar q_{f,r}^{\alpha}q_{f,s}^{\beta}
|0\rangle
\langle 0|
\bar q_{g,t}^{\gamma}q_{g,u}^{\delta}
|0\rangle
\notag\\
&\hspace{2.5cm}
-
\langle 0|
\bar q_{f,r}^{\alpha}q_{g,u}^{\delta}
|0\rangle
\langle 0|
\bar q_{g,t}^{\gamma}q_{f,s}^{\beta}
|0\rangle
\bigg],
\end{align}
where the first and second terms are the direct and exchange
contractions, respectively.

Lorentz, color, and flavor symmetries constrain the quark-bilinear
vacuum expectation value to
\begin{align}
\langle 0|
\bar q_{f,r}^{\alpha}q_{g,u}^{\delta}
|0\rangle
&=
\frac{\delta_{fg}\delta^{\alpha\delta}\delta_{ru}}
{4N_c}
\langle\bar q_f q_f\rangle_0 .
\end{align}
The color factors of the direct and exchange contractions are therefore
\begin{align}
\delta_{\alpha\beta}\delta_{\gamma\delta}
(\lambda^a)_{\alpha\beta}
(\lambda^a)_{\gamma\delta}
&=
\operatorname{Tr}(\lambda^a)
\operatorname{Tr}(\lambda^a)
=0,
\notag\\
\delta_{\alpha\delta}\delta_{\gamma\beta}
(\lambda^a)_{\alpha\beta}
(\lambda^a)_{\gamma\delta}
&=
\operatorname{Tr}(\lambda^a\lambda^a),
\end{align}
so that only the exchange contraction contributes.

Substituting the quark-bilinear expectation value into the exchange
term gives
\begin{align}
\langle\mathcal O_{4q}\rangle_0
&\simeq
-\frac{1}{(4N_c)^2}
\sum_f
\langle\bar q_f q_f\rangle_0^2
\sum_a
\operatorname{Tr}(\lambda^a\lambda^a)
\operatorname{Tr}(\gamma_\mu\gamma^\mu)
\notag\\
&=
-\frac{16}{9}
\sum_f
\langle\bar q_f q_f\rangle_0^2 ,
\end{align}
where $N_c=3$,
$\sum_a\operatorname{Tr}(\lambda^a\lambda^a)=16$, and
$\operatorname{Tr}(\gamma_\mu\gamma^\mu)=16$ have been used.

In symmetric nuclear matter, the linear-density contribution to the expectation value of a local operator $\mathcal O$ is written as~\cite{Drukarev:1990,Gubler:2019}
\begin{align}
\Delta\langle\mathcal O\rangle_\rho
&\equiv
\langle\mathcal O\rangle_\rho
-\langle\mathcal O\rangle_0
\simeq
\frac{\rho}{4}
\sum_{N=p,n}\sum_s
\frac{1}{2m_N}
\langle N(\mathbf 0,s)|
\mathcal O
|N(\mathbf 0,s)\rangle ,
\end{align}
where the factor $1/4$ averages over the four spin-isospin states of
the nucleon, while $1/(2m_N)$ follows from the relativistic normalization
\begin{align}
\langle N(\mathbf p’,s’)|N(\mathbf p,s)\rangle
&=
2E_p(2\pi)^3
\delta^{(3)}(\mathbf p’-\mathbf p)
\delta_{s’s}.
\end{align}
We denote the four-momentum of the external nucleon by $p^\mu=(m_N,\mathbf 0)$. 

For the nucleon matrix element, we insert a complete set of physical
intermediate states between the two quark bilinears. Baryon-number
conservation restricts the intermediate states to those with baryon
number one, and the nucleon is the lowest such state. Since all
physical states involved are color singlets, the direct contraction
vanishes as in the vacuum case, leaving only the exchange contribution,
\begin{align}
\Delta\langle\mathcal O_{4q}\rangle_\rho
\simeq
-\frac{\rho}{4}
\sum_{N=p,n}\sum_s
&\frac{1}{2m_N}
\sum_{f,g}
(\gamma_\mu)_{rs}
(\lambda^a)_{\alpha\beta}
(\gamma^\mu)_{tu}
(\lambda^a)_{\gamma\delta}
\notag\\
\times
\langle N(\mathbf 0,s)|
&\bar q_{f,r}^{\alpha}q_{g,u}^{\delta}
\left(\sum_H |H\rangle\langle H|\right)
\bar q_{g,t}^{\gamma}q_{f,s}^{\beta}
|N(\mathbf 0,s)\rangle .
\end{align}
Here, the sum over $H$ schematically represents the complete set, including the appropriate sums and phase-space integrations over intermediate states.

Compared with the vacuum case, the quark bilinears can have more general structures, depending on the
states retained in the complete-set insertion. We therefore first
project each quark bilinear onto its color-singlet component:
\begin{align}
\bar q^{\,\alpha}_{f,r}\,q^\delta_{g,u}
&\longrightarrow
\frac{\delta^{\alpha\delta}}{12}
\Bigl[
\delta_{ur}\,\bar q_f q_g
+(\gamma_5)_{ur}\,\bar q_f\gamma_5 q_g
\nonumber\\
&\qquad
+(\gamma_\nu)_{ur}\,\bar q_f\gamma^\nu q_g
+(\gamma_5\gamma_\nu)_{ur}\,
 \bar q_f\gamma^\nu\gamma_5 q_g
\nonumber\\
&\qquad
+\frac{1}{2}(\sigma_{\nu\rho})_{ur}\,
 \bar q_f\sigma^{\nu\rho}q_g
\Bigr].
\label{eq:bilinear_projection}
\end{align}
Among these Dirac channels, we retain the terms involving two pseudoscalar or two axial-vector bilinears associated with the pion-coupled contribution considered in Ref.~\cite{YeoEtAl2025}. The corresponding contribution is then
\begin{align}
\Delta\langle\mathcal O_{4q}^{\pi\pi}\rangle_\rho
&\simeq
\frac{\rho}{4}
\sum_{N=p,n}\sum_s
\frac{1}{2m_N}
\sum_{f,g}\notag\\
&\hspace{-1em}\times\bigg[
\frac{16}{9}
\langle N(\mathbf 0,s)|
\bar q_f\gamma_5 q_g
\left(\sum_H |H\rangle\langle H|\right)
\bar q_g\gamma_5 q_f
|N(\mathbf 0,s)\rangle
\notag\\
&\hspace{-2em}+
\frac{8}{9}
\langle N(\mathbf 0,s)|
\bar q_f\gamma^\mu\gamma_5 q_g
\left(\sum_H |H\rangle\langle H|\right)
\bar q_g\gamma_\mu\gamma_5 q_f
|N(\mathbf 0,s)\rangle
\bigg].
\end{align}

In the present calculation, we retain the one-nucleon part of this
complete set:
\begin{align}
\mathbf 1_{\mathrm{1N}}
&=
\sum_{N_2=p,n}\sum_{s_2}
\int_{|\mathbf q|<\Lambda}
\frac{d^3\mathbf q}{(2\pi)^3 2E_q}
|N_2(\mathbf q,s_2)\rangle
\langle N_2(\mathbf q,s_2)|,
\end{align}
where $q^\mu=(E_q,\mathbf q)$ with $E_q=\sqrt{m_N^2+\mathbf q^2}$.
The cutoff $\Lambda$ specifies the momentum region in which
one-nucleon intermediate-state saturation is assumed to describe the
low-energy part of the local four-quark matrix element. It is
therefore regarded as a hadronic separation scale rather than a
fundamental ultraviolet cutoff. The separation scale should be chosen
sufficiently large to capture the relevant low-energy hadronic
contribution, but not so large as to include momentum regions where
short-distance physics becomes important.

We next rewrite the flavor-summed pseudoscalar and axial-vector
contributions in the isospin basis. With $q=(u,d)^T$, we define
\begin{align}
P^{(0)}&\equiv\bar q i\gamma_5 q,
&
P^{(a)}&\equiv\bar q i\gamma_5\tau^a q,
\nonumber\\
A^{(0)\mu}&\equiv\bar q\gamma^\mu\gamma_5 q,
&
A^{(a)\mu}&\equiv\bar q\gamma^\mu\gamma_5\tau^a q.
\label{eq:isospin_bilinears}
\end{align}
The flavor sums can then be written as
\begin{align}
\sum_{f,g=u,d}
(\bar q_f\gamma_5 q_g)
(\bar q_g\gamma_5 q_f)
&=
-\frac{1}{2}
\left[
P^{(0)}P^{(0)}
+
P^{(a)}P^{(a)}
\right],
\\
\sum_{f,g=u,d}
(\bar q_f\gamma^\mu\gamma_5 q_g)
(\bar q_g\gamma_\mu\gamma_5 q_f)
&=
\frac{1}{2}
\left[
A^{(0)\mu}A^{(0)}_{\mu}
+
A^{(a)\mu}A^{(a)}_{\mu}
\right],
\end{align}
where a sum over $a=1,2,3$ is understood. The flavor sums include only $u$ and $d$ quarks for the pion-coupled contribution.

Inserting the one-nucleon projector between the two bilinears, we obtain the pseudoscalar and axial-vector contributions
\begin{align}
\Delta\langle\mathcal O_{4q,PP}^{\pi\pi}\rangle_\rho^{\mathrm{1N}}
&=
-\frac{2}{9}
\sum_{N=p,n}\sum_s
\frac{\rho}{2m_N}
\bigg[
\langle N(\mathbf 0,s)|P^{(0)}
\mathbf 1_{1N}
P^{(0)}|N(\mathbf 0,s)\rangle
\notag\\
&\hspace{2.5cm}
+
\langle N(\mathbf 0,s)|P^{(a)}
\mathbf 1_{1N}
P^{(a)}|N(\mathbf 0,s)\rangle
\bigg],\\
\Delta\langle\mathcal O_{4q,AA}^{\pi\pi}\rangle_\rho^{\mathrm{1N}}
&=
\frac{1}{9}
\sum_{N=p,n}\sum_s
\frac{\rho}{2m_N}
\bigg[
\langle N(\mathbf 0,s)|A^{(0)\mu}
\mathbf 1_{1N}
A^{(0)}_{\mu}|N(\mathbf 0,s)\rangle
\notag\\
&\hspace{2.5cm}
+
\langle N(\mathbf 0,s)|A^{(a)\mu}
\mathbf 1_{1N}
A^{(a)}_{\mu}|N(\mathbf 0,s)\rangle
\bigg].
\end{align}
The superscript $\mathrm{1N}$ denotes the contribution obtained by retaining only one-nucleon intermediate states. The nucleon matrix elements are evaluated in terms of momentum-dependent form factors in the following section.

\subsection{One-nucleon intermediate-state contributions}
\label{subsec:one_nucleon}
\subsubsection{Form factors}
\label{subsec:form_factors}

The nucleon matrix elements needed below are expressed in terms of
nucleon form factors.
In the isospin-symmetric limit, the isoscalar operators are diagonal
in nucleon isospin, whereas the isovector operators carry Pauli matrices between the initial and final nucleon isospin indices. Since the isoscalar and isovector operators have the same Lorentz transformation properties, their nucleon matrix elements can be decomposed in the same form, apart from these isospin factors. With
$\Delta^\mu=p^\mu-q^\mu$ and $Q^2=-\Delta^2$, we define the form factors as~\cite{FuchsScherer2003,AlexandrouEtAl2024}:
\begin{align}
& \langle N(p,s)|P^{(0)}|N_2(q,s_2)\rangle
\nonumber\\
& \qquad =
\delta_{NN_2}\,
\bar u_N(p,s)i\gamma_5u_{N_2}(q,s_2)\,
G_0^P(Q^2),
\nonumber\\
&\langle N(p,s)|P^{(a)}|N_2(q,s_2)\rangle
\nonumber\\
& \qquad =
(\tau^a)_{NN_2}\,
\bar u_N(p,s)i\gamma_5u_{N_2}(q,s_2)\,
G_1^P(Q^2),
\label{eq:ff_pseudoscalar}
\\
&\langle N(p,s)|A^{(0)\mu}|N_2(q,s_2)\rangle
\nonumber\\
& \qquad =
\delta_{NN_2}\,
\bar u_N(p,s)
\Gamma^\mu_{A,0}(p,q)
u_{N_2}(q,s_2),
\nonumber\\
&\langle N(p,s)|A^{(a)\mu}|N_2(q,s_2)\rangle
\nonumber\\
& \qquad =
(\tau^a)_{NN_2}\,
\bar u_N(p,s)
\Gamma^\mu_{A,1}(p,q)
u_{N_2}(q,s_2),
\label{eq:ff_axial}
\end{align}
with
\begin{align}
\Gamma^\mu_{A,I}(p,q)
&=
\gamma^\mu\gamma_5\,G^A_{1,I}(Q^2)
+
\frac{\Delta^\mu}{2m_N}\gamma_5\,G^A_{2,I}(Q^2),~I=0,1.
\label{eq:ff_axial_vertex}
\end{align}
Here, $G^A_{1,I}$, $G^A_{2,I}$, and $G_I^P$ denote the axial,
induced-pseudoscalar, and pseudoscalar form factors, respectively.
For the numerical analysis, we use the continuum-limit $z^3$ parameterizations of the isovector form factors given in
Ref.~\cite{AlexandrouEtAl2024}, with central values from the two-state fits.
The $z$ variable is defined as
\begin{align}
z(Q^2;t_0)
&=
\frac{
\sqrt{t_{\rm cut}+Q^2}-\sqrt{t_{\rm cut}+t_0}
}{
\sqrt{t_{\rm cut}+Q^2}+\sqrt{t_{\rm cut}+t_0}
},
\qquad
t_{\rm cut}=(3m_\pi)^2,
\label{eq:z_variable}
\end{align}
with $m_\pi=0.135~\mathrm{GeV}$.
The form factors used in the numerical evaluation are parameterized as
\begin{align}
G^A_{1,1}(Q^2)
&=
\sum_{k=0}^{3} a_k^{A_1} z^k(Q^2;0),
\nonumber\\
(Q^2+m_\pi^2)G^A_{2,1}(Q^2)
&=
\sum_{k=0}^{3} a_k^{A_2} z^k(Q^2;-m_\pi^2),
\nonumber\\
(Q^2+m_\pi^2)m_qG_1^P(Q^2)
&=
\frac{m_\pi^2}{4m_N}
\sum_{k=0}^{3} a_k^P z^k(Q^2;-m_\pi^2),
\label{eq:z_form_factors}
\end{align}
where the coefficients are
\begin{align}
\{a_k^{A_1}\}
&=
\{1.245,-1.19,-0.54,-0.13\},
\nonumber\\
\{a_k^{A_2}\}
&=
\{4.62,-3.0,-4.7,-0.1\}~\mathrm{GeV}^2,
\nonumber\\
\{a_k^P\}
&=
\{4.62,-2.2,-2.9,-1.2\}~\mathrm{GeV}^2.
\label{eq:z_coefficients}
\end{align}
Here, $m_N=0.938~\mathrm{GeV}$. For the pseudoscalar form factor, we use the scale-independent
$m_qG_1^P$ parameterization of Ref.~\cite{AlexandrouEtAl2024} and
evolve the light-quark mass used in that work from
$m_q^{\overline{\rm MS}}(\mu=2~\mathrm{GeV})=3.636~\mathrm{MeV}$
to $m_q(\mu=1~\mathrm{GeV})\simeq4.91~\mathrm{MeV}$ using the
scale-conversion factor $1.35$ adopted in
Ref.~\cite{NavasEtAl2024}.

The nonsinglet axial Ward
identity gives~\cite{AlexandrouEtAl2024}
\begin{align}
\partial_\mu A^{(a)\mu}
&=
2m_q P^{(a)},
\nonumber\\
G^A_{1,1}(Q^2)
-\frac{Q^2}{4m_N^2}G^A_{2,1}(Q^2)
&=
\frac{m_q}{m_N}G_1^P(Q^2).
\label{eq:ff_pcac_isovector}
\end{align}

For the light-isoscalar current, the Ward identity
contains the $U(1)_A$ anomaly and can be written as~\cite{LiangEtAl2018}
\begin{align}
\partial_\mu A^{(0)\mu}
=
2m_qP^{(0)}+\mathcal Q,~
\mathcal Q
\equiv
N_f\,\frac{\alpha_s}{4\pi}
G_{\mu\nu}^a\widetilde G^{a\mu\nu},~ N_f=2,
\label{eq:ff_isoscalar_awi}
\end{align}
where
$\widetilde G^{a\mu\nu}
=\frac{1}{2}\epsilon^{\mu\nu\alpha\beta}G_{\alpha\beta}^a$.
Defining the anomaly form factor by
\begin{align}
&\langle N(p,s)|\mathcal Q|N_2(q,s_2)\rangle
\nonumber\\&=
2m_N\delta_{NN_2}\,
\bar u_N(p,s)i\gamma_5u_{N_2}(q,s_2)\,
G_{\rm anom}(Q^2),
\label{eq:ff_anomaly_form_factor}
\end{align}
the anomalous Ward identity becomes
\begin{align}
G^A_{1,0}(Q^2)
-\frac{Q^2}{4m_N^2}G^A_{2,0}(Q^2)
&=
\frac{m_q}{m_N}G_0^P(Q^2)
+G_{\rm anom}(Q^2).
\label{eq:ff_pcac_isoscalar}
\end{align}

In the forward limit,
Eqs.~\eqref{eq:ff_pcac_isovector} and
\eqref{eq:ff_pcac_isoscalar} give
\begin{align}
\frac{G_0^P(0)}{G_1^P(0)}
&=
\frac{
G^A_{1,0}(0)-G_{\rm anom}(0)
}{
G^A_{1,1}(0)
}.
\label{eq:ff_isoscalar_forward}
\end{align}
Denoting the axial charge associated with flavor $f$ by $\Delta f$~\cite{Green2017}, one has $G^A_{1,0}(0)=\Delta u+\Delta d$ and $G^A_{1,1}(0)=\Delta u-\Delta d=g_A$.

\subsubsection{Pseudoscalar channel}

For an external nucleon at rest, the spacelike
momentum transfer satisfies $Q^2=2m_N(E_q-m_N)\geq 0$.

With the pseudoscalar form-factor parameterizations in
Eq.~\eqref{eq:ff_pseudoscalar}, the isoscalar contraction gives unity, whereas
the isovector contraction gives
\begin{align}
\sum_{a=1}^3\sum_{N_2}
(\tau^a)_{NN_2}(\tau^a)_{N_2N}
&=
3.
\label{eq:isovector_factor}
\end{align}
The corresponding spinor contraction is
\begin{align}
\sum_{s,s_2}
&
\bar u_N(p,s)i\gamma_5u_{N_2}(q,s_2)
\bar u_{N_2}(q,s_2)i\gamma_5u_N(p,s)
\nonumber\\
&=
4m_N(E_q-m_N).
\label{eq:pseudoscalar_spinor_trace}
\end{align}
It is convenient to introduce the functional
\begin{align}
I_P[F]
&\equiv
\int_{|\boldsymbol q|<\Lambda}
\frac{d^3\boldsymbol q}{(2\pi)^3\,2E_q}\,
F(Q^2)(E_q-m_N).
\label{eq:IP_def}
\end{align}
After carrying out the spin and isospin sums, we find the one-nucleon pseudoscalar contribution
\begin{align}
\Delta\langle \mathcal O_{4q,PP}^{\pi\pi}\rangle_\rho^{\mathrm{1N}}
&=
-\frac{8\rho}{9}\,
I_P
\left[
\left(G_0^P\right)^2
+
3\left(G_1^P\right)^2
\right].
\label{eq:pseudoscalar_projected_ONI}
\end{align}

\subsubsection{Axial-vector channel}

Employing the axial form-factor parameterizations in
Eqs.~\eqref{eq:ff_axial} and \eqref{eq:ff_axial_vertex}, we retain
both the axial and induced-pseudoscalar structures. The sign of the
induced-pseudoscalar term changes in the reversed matrix element
because the momentum transfer changes from
$\Delta^\mu=p^\mu-q^\mu$ to $-\Delta^\mu$, while $Q^2$ remains
unchanged.

For each isospin channel, we define the axial kernel
\begin{align}
&K^A_I(Q^2)\nonumber\\
&\equiv
\frac{1}{2m_N}
\operatorname{Tr}
\left[
(\slashed p+m_N)
\Gamma^\mu_{A,I}(p,q)
(\slashed q+m_N)
\Gamma_{A,I,\mu}(q,p)
\right].
\label{eq:axial_kernel_def}
\end{align}
Using $p^2=q^2=m_N^2$ and $\Delta^2=-Q^2$, we have
\begin{align}
K^A_I(Q^2)
&=
-\left(
8m_N+\frac{2Q^2}{m_N}
\right)
\left[G^A_{1,I}(Q^2)\right]^2
\nonumber\\
&\qquad
-4m_N
\left[
G^A_{1,I}(Q^2)
-\frac{Q^2}{4m_N^2}G^A_{2,I}(Q^2)
\right]^2.
\label{eq:axial_kernel_square}
\end{align}
Hence $K^{A}_{I}(Q^2)\leq0$ for $Q^2\geq0$, with $K^{A}_{I}(0)=-12m_N [G^{A}_{1,I}(0)]^2$.
We also define
\begin{align}
I_A[F]
&\equiv
\int_{|\boldsymbol q|<\Lambda}
\frac{d^3\boldsymbol q}{(2\pi)^3\,2E_q}\,
F(Q^2).
\label{eq:IA_def}
\end{align}
The spin and isospin sums give the one-nucleon axial-vector contribution
\begin{align}
\Delta\langle \mathcal O_{4q,AA}^{\pi\pi}\rangle_\rho^{\mathrm{1N}}
&=
\frac{2\rho}{9}\,
I_A
\left[
K_0^A+3K_1^A
\right].
\label{eq:axial_projected_ONI}
\end{align}

Combining Eqs.~\eqref{eq:pseudoscalar_projected_ONI} and
\eqref{eq:axial_projected_ONI} gives the one-nucleon contribution to
the pion-coupled four-quark condensate:
\begin{align}
\Delta
\left\langle
\mathcal O_{4q}^{\pi\pi}
\right\rangle_\rho^{\mathrm{1N}}
&=
\Delta\langle \mathcal O_{4q,PP}^{\pi\pi}\rangle_\rho^{\mathrm{1N}}
+
\Delta\langle \mathcal O_{4q,AA}^{\pi\pi}\rangle_\rho^{\mathrm{1N}}
\nonumber\\
&=
-\frac{8\rho}{9}\,
I_P
\left[
\left(G_0^P\right)^2
+
3\left(G_1^P\right)^2
\right]
\nonumber\\
&\qquad
+\frac{2\rho}{9}\,
I_A
\left[
K_0^A+3K_1^A
\right].
\label{eq:projected_ONI_total}
\end{align}
When only the isovector intermediate-state contribution is retained,
this reduces to
\begin{align}
\Delta
\left\langle
\mathcal O_{4q}^{\pi\pi}
\right\rangle_\rho^{\mathrm{1N},\,I=1}
&=
-\frac{8\rho}{3}\,
I_P
\left[
\left(G_1^P\right)^2
\right]
+\frac{2\rho}{3}\,
I_A
\left[
K_1^A
\right].
\label{eq:projected_ONI_isovector}
\end{align}
For $\Lambda=1~\mathrm{GeV}$, our numerical evaluation gives
\begin{align}
I_P\left[\left(G_1^P\right)^2\right]
&=0.128~\mathrm{GeV}^3,
&
I_A\left[K_1^A\right]
&=-0.0394~\mathrm{GeV}^3.
\end{align}

The same method can be applied to
$\mathcal O_{4q,\mu\nu}^{\mathrm{spin}\text{-}2}$,
the symmetric-traceless counterpart of $\mathcal O_{4q}$. The PP term is proportional to $g_{\mu\nu}$ and is therefore removed by this projection, leaving only the AA term. The pion-coupled spin-2 contribution is approximately $2\%$ of its scalar counterpart. Including it in the sum-rule analysis also changes the mass shift by only about $2\%$. For convenience, we omit this term from the following analysis.

\subsubsection{Estimates of isoscalar contributions}
\label{isoscalar}

Our numerical analysis uses only the isovector form factors, since a complete set of continuum-limit isoscalar form factors is not yet available for a finite-$Q^2$ analysis~\cite{Green2017,AlexandrouEtAl2021}. Recent results from Ref.~\cite{BaroneEtAl2026} allow $G^A_{1,0}$ to be constructed in the continuum limit, whereas corresponding results for $G^A_{2,0}$ and $G^P_0$ are still lacking. The forward form-factor normalizations nevertheless provide
an initial estimate of the relative sizes of the isoscalar and isovector terms. For an illustrative estimate in the pseudoscalar channel, we neglect the $U(1)_A$ anomaly contribution to the
singlet Ward identity, as its magnitude is not well known.
This gives
\begin{align}
\frac{1}{3}
\left[
\frac{G_0^P(0)}{G_1^P(0)}
\right]^2
&\simeq
\frac{1}{3}
\left[
\frac{G_{1,0}^A(0)}{G_{1,1}^A(0)}
\right]^2
=
\frac{1}{3}
\left[
\frac{\Delta u+\Delta d}{g_A}
\right]^2
\sim 2\text{--}5\%.
\label{eq:isoscalar_pseudoscalar_estimate}
\end{align}
Here, we use the typical range
$\Delta u+\Delta d\simeq0.3$--$0.5$~\cite{LiangEtAl2018,Alexandrou2020,Lin2018,Green2017}
together with $g_A\simeq1.27$. For the axial-vector channel, the forward kernel depends only on the axial form factor $G_{1,I}^A(0)$, yielding
\begin{align}
\frac{K_0^A(0)}{3K_1^A(0)}
&=
\frac{1}{3}
\left[
\frac{G_{1,0}^A(0)}{G_{1,1}^A(0)}
\right]^2
=
\frac{1}{3}
\left[
\frac{\Delta u+\Delta d}{g_A}
\right]^2
\sim 2\text{--}5\%.
\label{eq:isoscalar_axial_estimate}
\end{align}
At finite $Q^2$, however, both $G_{1,I}^A$ and $G_{2,I}^A$ contribute to the axial kernel, so that the forward-limit estimate alone does not determine the integrated contribution.

Since the pseudoscalar term is quadratic in $G_0^P$ and $K_0^A(Q^2)\leq0$, the isoscalar terms have the same overall signs as their isovector counterparts and would therefore further enhance the magnitude of the one-nucleon result. Although the isoscalar axial form factors in Ref.~\cite{AlexandrouEtAl2021} were obtained at a single lattice spacing, an estimate based on them gives an isoscalar axial-vector contribution $I_A[K_0^A]$ of only a few percent of its isovector counterpart $3I_A[K_1^A]$ in magnitude.

\section{OPE and sum rule setup}
\label{qsr}

We use the $J/\psi$ QCD sum rule of
Ref.~\cite{YeoEtAl2025}, with the vector current
$j_\mu=\bar c\gamma_\mu c$. At vanishing spatial momentum, the
dimensionless correlator is defined as
$\widetilde\Pi(q^2)=-\Pi^\mu_{\ \mu}(q)/(3q^2)$~\cite{Morita2010}. After performing the OPE up to dimension six and applying the Borel transform, we get
\begin{align}
\mathcal M_{\mathrm{OPE}}^V(\nu)
&=
e^{-\nu}A^V(\nu)
\Biggl[
1+a^V(\nu)\alpha_s(\nu)
+b^V(\nu)\phi_b\nonumber\\
&+s^V(\nu)\phi_c
+\left(
2s^V(\nu)+\frac{2}{3}t^V(\nu)
\right)\phi_d
+\widetilde y^V(\nu)\phi_e
\Biggr],
\label{eq:borel_OPE}
\end{align}
where $\nu=4m_c^2/M^2$. We use the Wilson coefficients from Refs.~\cite{nikolaev1983,KimLee2001,Morita2010,YeoEtAl2025} and the definitions of $\phi_i$ given in Ref.~\cite{YeoEtAl2025}. The spin-zero dimension-six term is
\begin{align}
\phi_d
&\equiv
\frac{4\pi^2}
{3\times1080(4m_c^2)^3}
\left\langle
-\alpha_s^2
\left(
\bar q_i\gamma_\mu\lambda^a q_i
\right)^2
\right\rangle.
\label{eq:phi_d_def}
\end{align}
For the strong coupling, we distinguish between the perturbative and
nonperturbative sectors of the OPE. The perturbative radiative correction uses $\alpha_s(\nu)$ obtained by solving the running-coupling equation with the input
$\alpha_s(8m_c^2)=0.21$. For the
nonperturbative condensate contributions, we use the fixed infrared
value $\alpha_{s,\mathrm{IR}}=0.7$~\cite{nikolaev1983,KimLee2001,Deur2016} for both the vacuum and
density-dependent parts.

In this analysis, the density-dependent pion-coupled part of $\phi_d$ is evaluated using the one-nucleon contribution obtained in Sec.~\ref{subsec:one_nucleon}, which gives
\begin{align}
\Delta\phi_d^{\mathrm{1N}}(\rho)
&=
-\frac{4\pi^2\alpha_{s,\mathrm{IR}}^2}
{3\times1080(4m_c^2)^3}
\Delta
\left\langle
\mathcal O_{4q}^{\pi\pi}
\right\rangle_\rho^{\mathrm{1N}},
\label{eq:phi_d_ONI}
\end{align}
where we use Eq.~\eqref{eq:projected_ONI_isovector} for the expectation value. This term replaces the factorized density-dependent four-quark contribution used in Ref.~\cite{YeoEtAl2025}. The vacuum part of the OPE and all remaining condensate inputs are kept unchanged.

On the phenomenological side, we use the same pole-plus-continuum
ansatz as in Ref.~\cite{YeoEtAl2025}. We define the
continuum-subtracted Borel moment as
\begin{align}
\overline{\mathcal M}^V(M^2,s_0)
&=
\mathcal M_{\mathrm{OPE}}^V(M^2)
-
\int_{s_0}^{\infty}
ds\,e^{-s/M^2}\rho_{\mathrm{pert}}^V(s).
\label{eq:continuum_subtracted_moment}
\end{align}
The $J/\psi$ mass is then extracted from
\begin{align}
m_{J/\psi}(M^2,s_0)
&=
\left[
-\frac{
\partial\overline{\mathcal M}^V(M^2,s_0)/
\partial(1/M^2)
}{
\overline{\mathcal M}^V(M^2,s_0)
}
\right]^{1/2}.
\label{eq:Jpsi_mass}
\end{align}
We impose the same Borel-window criteria and continuum-threshold
optimization as in Ref.~\cite{YeoEtAl2025}.

\section{Results}
\label{qsr_results}

We evaluate the density-dependent four-quark contribution at the
saturation density $\rho_0=0.17~\mathrm{fm}^{-3}$. Because the
isoscalar form factors are not sufficiently constrained, the
numerical analysis retains only the isovector part. At $\Lambda=1~\mathrm{GeV}$, we find 
\begin{align}
\Delta
\left\langle
\mathcal O_{4q}^{\pi\pi}
\right\rangle_{\rho_0}^{\mathrm{1N},\,I=1}
&=
-47.97\times10^{-5}~\mathrm{GeV}^6.
\label{eq:ONI_numerical_result}
\end{align}
The result is dominated by the pseudoscalar channel. Among the three form factors considered here, $G_1^P$ and $G_{2,1}^A$ exhibit a pion-pole structure~\cite{AlexandrouEtAl2024}. Schematically, in the pseudoscalar channel, this structure combines with the kinematic factor to give
$(E_q-m_N)/(Q^2+m_\pi^2)^2
=Q^2/(2m_N(Q^2+m_\pi^2)^2)$,
which is maximal at $Q^2=m_\pi^2$. By contrast, the terms involving $G_{2,1}^A$ in the axial kernel behave as $Q^2/(Q^2+m_\pi^2)$ and its square, without a similar low-$Q^2$ enhancement. Our result is approximately $5.8$ times larger in magnitude than the factorized linear-density estimate $\Delta\langle\mathcal O_{4q}^{\pi\pi}\rangle_{\rho_0}^{\mathrm{fac}}=-8.21\times10^{-5}~\mathrm{GeV}^6$ obtained in Ref.~\cite{YeoEtAl2025}.

Taking $\Lambda=1$ GeV as a representative hadronic separation scale, we substitute Eq.~\eqref{eq:ONI_numerical_result} into Eq.~\eqref{eq:phi_d_ONI} and apply the sum-rule analysis. The resulting mass shift is
\begin{align}
\Delta m_{J/\psi}^{\mathrm{1N}}
&=
-1.75~\mathrm{MeV}
\qquad
(\Lambda=1~\mathrm{GeV}).
\label{eq:Jpsi_mass_shift_ONI}
\end{align}
Varying the cutoff to $\Lambda=0.8$ and $1.2~\mathrm{GeV}$ yields mass shifts of $-1.42$ and $-2.01~\mathrm{MeV}$, respectively. The lattice-based estimate of the mass shift is
\begin{align}
\Delta m_{J/\psi}^{\mathrm{TPE}}
&\simeq
-5~\mathrm{MeV}.
\label{eq:lattice_TPE_mass_shift}
\end{align}
This value was obtained in Ref.~\cite{YeoEtAl2025} by integrating the lattice-fitted two-pion-exchange potential of Ref.~\cite{LyuEtAl2025} over the long-distance region
$1.0<r<1.8~\mathrm{fm}$ at saturation density. The one-nucleon evaluation substantially increases the attractive four-quark contribution relative to the previous treatment and brings the QCD sum-rule result closer to this lattice-based reference value without requiring the large values of $\kappa_{\Delta\rho}$ and $\alpha_{s,\mathrm{IR}}$ adopted in Ref.~\cite{YeoEtAl2025}.

Additional hadronic intermediate states may provide further contributions to the four-quark matrix element. Higher-order corrections to the four-quark Wilson coefficient may also be important for the long-distance two-pion exchange contribution. Radiative corrections to the dimension-four gluon-condensate contribution have been studied in heavy-quark current correlators and charmonium sum rules~\cite{Broadhurst:1994qj,Ioffe:2002be}. For the dimension-six four-quark Wilson coefficient considered here, however, higher-order corrections remain to be calculated and could modify the magnitude of the mass shift.

\section{Summary}
\label{summary}

We investigated the density-dependent pseudoscalar and axial-vector four-quark condensates relevant to the long-range $N$--$J/\psi$ interaction by inserting one-nucleon intermediate states. Using momentum-dependent nucleon form factors, we derived the isoscalar and isovector contributions and performed the numerical analysis in
the isovector channel. At $\Lambda=1~\mathrm{GeV}$, the isovector one-nucleon contribution is larger in magnitude than the corresponding factorized estimate by a factor of approximately $5.8$. The enhancement is dominated by the pseudoscalar channel. In the forward limit, the isoscalar contributions are estimated to be
only a few percent of their isovector counterparts, with the pseudoscalar ratio obtained by neglecting the anomaly contribution. They have the same overall signs and would therefore further increase the magnitude of the one-nucleon result. Consistent with this expectation, an estimate using isoscalar axial-vector form factors from a single lattice ensemble gives a correction of the same sign and only a few percent in magnitude relative to the isovector contribution.

Incorporating the one-nucleon result into the $J/\psi$ QCD sum rule
gives an attractive mass shift of
$\Delta m_{J/\psi}=-1.75~\mathrm{MeV}$. This value is of the same order
as the lattice-based two-pion exchange estimate of approximately
$-5~\mathrm{MeV}$. Additional hadronic intermediate states and higher-order corrections to the four-quark Wilson coefficient may affect the magnitude of the mass shift. Despite these uncertainties, this work shows that the explicit one-nucleon intermediate-state contribution can yield a substantially larger in magnitude four-quark condensate estimate than the previous ground-state saturation treatment and improve the connection between the OPE description and the long-distance $N$--$J/\psi$ interaction.

Determining the density dependence of four-quark condensates is a challenging problem. A commonly used treatment, referred to as the GSA, is simply to extend the vacuum saturation hypothesis to finite density and retain the linear-density contribution. Such an approximation might be reliable for chiral symmetry breaking four-quark operators~\cite{KimLee2021,KimLee2022}. 
Although we have only included the form factors that couple to the pions, by incorporating other form factors associated with Eq.~\eqref{eq:bilinear_projection}, the intermediate-state approach that we adopted here allows us to study any four-quark operators in medium to linear order in density by evaluating the relevant nucleon form factors directly.  This approach makes the hadronic content of four-quark matrix elements more explicit and can be applied to other hadronic channels.

\section*{Acknowledgements}
This work was supported by the Korea National Research Foundation under grant No. RS-2023-NR077232. 
P.~G.~is supported by the JSPS Grant-in-Aid for Scientific Research (S) under Grant No.~JP25H00400 and 
the JSPS Grant-in-Aid for Scientific Research (C) under Grant No.~JP26K07113.

\section*{Data availability}
All numerical results generated in this study are provided in this article. The input parameters are specified in the text or taken from the cited references.

\bibliographystyle{elsarticle-num}
\bibliography{refs}

@article{HatsudaLee1992,
  author  = {Hatsuda, Tetsuo and Lee, Su Houng},
  title   = {{QCD} sum rules for vector mesons in the nuclear medium},
  journal = {Phys. Rev. C},
  volume  = {46},
  pages   = {R34--R38},
  year    = {1992},
  doi     = {10.1103/PhysRevC.46.R34}
}

@article{SVZ1979,
  author  = {Shifman, Mikhail A. and Vainshtein, Arkady I. and
             Zakharov, Valentin I.},
  title   = {{QCD} and resonance physics: Theoretical foundations},
  journal = {Nucl. Phys. B},
  volume  = {147},
  pages   = {385--447},
  year    = {1979},
  doi     = {10.1016/0550-3213(79)90022-1}
}

@article{ReindersEtAl1985,
  author  = {Reinders, Leonard J. and Rubinstein, Howard and Yazaki, Shunji},
  title   = {Hadron properties from {QCD} sum rules},
  journal = {Phys. Rept.},
  volume  = {127},
  pages   = {1--97},
  year    = {1985},
  doi     = {10.1016/0370-1573(85)90065-1}
}

@article{BrodskyEtAl1990,
  author  = {Brodsky, Stanley J. and Schmidt, Ivan A. and
             de T{\'e}ramond, Guy F.},
  title   = {Nuclear-bound quarkonium},
  journal = {Phys. Rev. Lett.},
  volume  = {64},
  pages   = {1011--1014},
  year    = {1990},
  doi     = {10.1103/PhysRevLett.64.1011}
}

@article{KlinglEtAl1999,
  author        = {Klingl, Frank and Kim, Sungsik and Lee, Su Houng and
                   Morath, Philippe and Weise, Wolfram},
  title         = {{$J/\psi$ and $\eta_c$} in the nuclear medium:
                   {QCD} sum rule approach},
  journal       = {Phys. Rev. Lett.},
  volume        = {82},
  pages         = {3396--3399},
  year          = {1999},
  doi           = {10.1103/PhysRevLett.82.3396},
  eprint        = {nucl-th/9811070},
  archivePrefix = {arXiv},
  note          = {Erratum: Phys. Rev. Lett. 83, 4224 (1999)}
}

@article{LeeKo2003,
  author        = {Lee, Su Houng and Ko, Che Ming},
  title         = {Charmonium mass in nuclear matter},
  journal       = {Phys. Rev. C},
  volume        = {67},
  pages         = {038202},
  year          = {2003},
  doi           = {10.1103/PhysRevC.67.038202},
  eprint        = {nucl-th/0208003},
  archivePrefix = {arXiv}
}

@article{LyuEtAl2022,
  author        = {Lyu, Yan and Doi, Takumi and Hatsuda, Tetsuo and
                   Ikeda, Yoichi and Meng, Jie and Sasaki, Kenji and
                   Sugiura, Takuya},
  title         = {Attractive {$N$--$\phi$} interaction and two-pion
                   tail from lattice {QCD} near physical point},
  journal       = {Phys. Rev. D},
  volume        = {106},
  pages         = {074507},
  year          = {2022},
  doi           = {10.1103/PhysRevD.106.074507},
  eprint        = {2205.10544},
  archivePrefix = {arXiv},
  primaryClass  = {hep-lat}
}

@article{LyuEtAl2025,
  author        = {Lyu, Yan and Doi, Takumi and Hatsuda, Tetsuo and
                   Sugiura, Takuya},
  title         = {Nucleon-charmonium interactions from lattice {QCD}},
  journal       = {Phys. Lett. B},
  volume        = {860},
  pages         = {139178},
  year          = {2025},
  doi           = {10.1016/j.physletb.2024.139178},
  eprint        = {2410.22755},
  archivePrefix = {arXiv},
  primaryClass  = {hep-lat}
}

@article{WuEtAl2025,
  author        = {Wu, Bing and Dong, Xiang-Kun and Du, Meng-Lin and
                   Guo, Feng-Kun and Zou, Bing-Song},
  title         = {Deciphering the mechanism of {$J/\psi$}-nucleon
                   scattering},
  journal       = {Fundamental Research},
  volume        = {5},
  number        = {6},
  pages         = {2530--2536},
  year          = {2025},
  doi           = {10.1016/j.fmre.2025.07.005},
  eprint        = {2410.19526},
  archivePrefix = {arXiv},
  primaryClass  = {hep-ph}
}

@article{Hayashigaki1999,
  author        = {Hayashigaki, Arata},
  title         = {{$J/\psi$}-nucleon scattering length and in-medium
                   mass shift of {$J/\psi$} in {QCD} sum rule analysis},
  journal       = {Prog. Theor. Phys.},
  volume        = {101},
  pages         = {923--935},
  year          = {1999},
  doi           = {10.1143/PTP.101.923},
  eprint        = {nucl-th/9811092},
  archivePrefix = {arXiv}
}

@article{KimLee2001,
  author        = {Kim, Sungsik and Lee, Su Houng},
  title         = {{QCD} sum rules for {$J/\psi$} in the nuclear medium:
                   Contributions from dimension-six operators},
  journal       = {Nucl. Phys. A},
  volume        = {679},
  pages         = {517--548},
  year          = {2001},
  eprint        = {nucl-th/0002002},
  archivePrefix = {arXiv}
}

@article{YeoEtAl2025,
  author        = {Yeo, Seokwoo and Park, In Woo and Lee, Su Houng},
  title         = {Long-range {$N$--$J/\psi$} interaction from an
                   operator product expansion perspective},
  journal       = {Phys. Rev. D},
  volume        = {112},
  pages         = {094044},
  year          = {2025},
  doi           = {10.1103/n7z5-9fd1},
  eprint        = {2508.04980},
  archivePrefix = {arXiv},
  primaryClass  = {hep-ph}
}

@article{KimLee2021,
  author        = {Kim, Jisu and Lee, Su Houng},
  title         = {Vector meson mass in the chiral symmetry restored
                   vacuum},
  journal       = {Phys. Rev. D},
  volume        = {103},
  pages         = {L051501},
  year          = {2021},
  doi           = {10.1103/PhysRevD.103.L051501},
  eprint        = {2012.06463},
  archivePrefix = {arXiv},
  primaryClass  = {hep-ph}
}

@article{KimLee2022,
  author        = {Kim, Jisu and Lee, Su Houng},
  title         = {Masses of hadrons in the chiral symmetry restored
                   vacuum},
  journal       = {Phys. Rev. D},
  volume        = {105},
  pages         = {014014},
  year          = {2022},
  doi           = {10.1103/PhysRevD.105.014014},
  eprint        = {2109.12791},
  archivePrefix = {arXiv},
  primaryClass  = {hep-ph}
}

@article{AlexandrouEtAl2024,
  author        = {Alexandrou, Constantia and Bacchio, Simone and
                   Constantinou, Martha and Finkenrath, Jacob and
                   Frezzotti, Roberto and Kostrzewa, Bartosz and
                   Koutsou, Giannis and Spanoudes, Gregoris and
                   Urbach, Carsten},
  title         = {Nucleon axial and pseudoscalar form factors using
                   twisted-mass fermion ensembles at the physical point},
  journal       = {Phys. Rev. D},
  volume        = {109},
  pages         = {034503},
  year          = {2024},
  doi           = {10.1103/PhysRevD.109.034503},
  eprint        = {2309.05774},
  archivePrefix = {arXiv},
  primaryClass  = {hep-lat}
}

@article{nikolaev1983,
title = {Vacuum corrections to QCD charmonium sum rules: Basic formalism and O(G3) results},
journal = {Nuclear Physics B},
volume = {213},
number = {2},
pages = {285-304},
year = {1983},
issn = {0550-3213},
doi = {https://doi.org/10.1016/0550-3213(83)90513-8},
author = {S.N. Nikolaev and A.V. Radyushkin}
}

@article{Alexandrou2020,
  author        = {Alexandrou, Constantia and Bacchio, Simone
                   and Constantinou, Martha and Finkenrath, Jacob
                   and Hadjiyiannakou, Kyriakos and Jansen, Karl
                   and Koutsou, Giannis and Panagopoulos, Haralambos
                   and Spanoudes, Gregoris},
  title         = {Complete flavor decomposition of the spin and momentum
                   fraction of the proton using lattice {QCD} simulations
                   at physical pion mass},
  journal       = {Phys. Rev. D},
  volume        = {101},
  pages         = {094513},
  year          = {2020},
  doi           = {10.1103/PhysRevD.101.094513},
  eprint        = {2003.08486},
  archivePrefix = {arXiv},
  primaryClass  = {hep-lat}
}

@article{Lin2018,
  author        = {Lin, Huey-Wen and Gupta, Rajan and Yoon, Boram
                   and Jang, Yong-Chull and Bhattacharya, Tanmoy},
  title         = {Quark contribution to the proton spin from
                   {$2+1+1$}-flavor lattice {QCD}},
  journal       = {Phys. Rev. D},
  volume        = {98},
  pages         = {094512},
  year          = {2018},
  doi           = {10.1103/PhysRevD.98.094512},
  eprint        = {1806.10604},
  archivePrefix = {arXiv},
  primaryClass  = {hep-lat}
}

@article{Green2017,
  author        = {Green, Jeremy and Hasan, Nesreen and Meinel, Stefan
                   and Engelhardt, Michael and Krieg, Stefan
                   and Laeuchli, Jesse and Negele, John
                   and Orginos, Kostas and Pochinsky, Andrew
                   and Syritsyn, Sergey},
  title         = {Up, down, and strange nucleon axial form factors
                   from lattice {QCD}},
  journal       = {Phys. Rev. D},
  volume        = {95},
  pages         = {114502},
  year          = {2017},
  doi           = {10.1103/PhysRevD.95.114502},
  eprint        = {1703.06703},
  archivePrefix = {arXiv},
  primaryClass  = {hep-lat}
}

@article{Morita2010,
   title={Heavy quarkonium correlators at finite temperature: QCD sum rule approach},
   volume={82},
   ISSN={1550-2368},
   DOI={10.1103/physrevd.82.054008},
   number={5},
   journal={Physical Review D},
   publisher={American Physical Society (APS)},
   author={Morita, Kenji and Lee, Su Houng},
   year={2010},
   month=sep }

@article{Deur2016,
   title={The QCD running coupling},
   volume={90},
   ISSN={0146-6410},
   DOI={10.1016/j.ppnp.2016.04.003},
   journal={Progress in Particle and Nuclear Physics},
   publisher={Elsevier BV},
   author={Deur, Alexandre and Brodsky, Stanley J. and de Téramond, Guy F.},
   year={2016},
   month=sep, pages={1–74} }

@article{Leupold2005,
  author  = {Leupold, Stefan},
  title   = {Factorization and non-factorization of in-medium four-quark condensates},
  journal = {Phys. Lett. B},
  volume  = {616},
  pages   = {203--207},
  year    = {2005},
  doi     = {10.1016/j.physletb.2005.04.046}
}

@article{FuchsScherer2003,
  author        = {Fuchs, T. and Scherer, S.},
  title         = {Pion electroproduction, PCAC, chiral Ward identities,
                   and the axial form factor revisited},
  journal       = {Phys. Rev. C},
  volume        = {68},
  pages         = {055501},
  year          = {2003},
  doi           = {10.1103/PhysRevC.68.055501},
  eprint        = {nucl-th/0303002},
  archivePrefix = {arXiv}
}

@article{LiangEtAl2018,
  author        = {Liang, Jian and Yang, Yi-Bo and Draper, Terrence
                   and Gong, Ming and Liu, Keh-Fei},
  title         = {Quark spins and anomalous Ward identity},
  journal       = {Phys. Rev. D},
  volume        = {98},
  pages         = {074505},
  year          = {2018},
  doi           = {10.1103/PhysRevD.98.074505},
  eprint        = {1806.08366},
  archivePrefix = {arXiv},
  primaryClass  = {hep-ph}
}

@article{NavasEtAl2024,
  author        = {Navas, S. and others},
  collaboration = {Particle Data Group},
  title         = {Review of Particle Physics},
  journal       = {Phys. Rev. D},
  volume        = {110},
  pages         = {030001},
  year          = {2024},
  doi           = {10.1103/PhysRevD.110.030001}
}

@article{Hatsuda2026,
  author  = {Hatsuda, Tetsuo},
  title   = {Two-pion exchange paradigm in hadron-hadron interactions from lattice QCD},
  journal = {Journal of Subatomic Particles and Cosmology},
  volume  = {5},
  pages   = {100341},
  year    = {2026},
  doi     = {10.1016/j.jspc.2026.100341}
}

@article{AlexandrouEtAl2021,
  author        = {Alexandrou, Constantia and Bacchio, Simone
                   and Constantinou, Martha and Hadjiyiannakou, Kyriakos
                   and Jansen, Karl and Koutsou, Giannis},
  title         = {Quark flavor decomposition of the nucleon axial form factors},
  journal       = {Phys. Rev. D},
  volume        = {104},
  pages         = {074503},
  year          = {2021},
  doi           = {10.1103/PhysRevD.104.074503},
  eprint        = {2106.13468},
  archivePrefix = {arXiv},
  primaryClass  = {hep-lat}
}

@article{BaroneEtAl2026,
  author        = {Barone, Alessandro and Djukanovic, Dalibor
                   and von Hippel, Georg and Meyer, Harvey B.
                   and Ottnad, Konstantin and Wittig, Hartmut},
  title         = {The strange and flavor-singlet axial form factors
                   of the nucleon from lattice QCD},
  year          = {2026},
  eprint        = {2605.06559},
  archivePrefix = {arXiv},
  primaryClass  = {hep-lat}
}

@article{Peskin1979,
author  = {Peskin, Michael E.},
title   = {Short-distance analysis for heavy-quark systems. I. Diagrammatics},
journal = {Nucl. Phys. B},
volume  = {156},
pages   = {365–390},
year    = {1979},
doi     = {10.1016/0550-3213(79)90199-8}
}

@article{FujiiKharzeev1999,
author        = {Fujii, Hirotsugu and Kharzeev, Dmitri},
title         = {Long-range forces of {QCD}},
journal       = {Phys. Rev. D},
volume        = {60},
pages         = {114039},
year          = {1999},
doi           = {10.1103/PhysRevD.60.114039},
eprint        = {hep-ph/9903495},
archivePrefix = {arXiv}
}

@article{Broadhurst:1994qj,
    author = "Broadhurst, D. J. and Baikov, P. A. and Ilyin, V. A. and Fleischer, J. and Tarasov, O. V. and Smirnov, V. A.",
    title = "{Two Loop Gluon Condensate Contributions to Heavy Quark Current Correlators: Exact Results and Approximations}",
    eprint = "hep-ph/9403274",
    archivePrefix = "arXiv",
    journal = "Phys. Lett. B",
    volume = "329",
    pages = "103--110",
    year = "1994",
    doi = "10.1016/0370-2693(94)90524-X"
}

@article{Ioffe:2002be,
    author = "Ioffe, B. L. and Zyablyuk, K. N.",
    title = "{Gluon Condensate in Charmonium Sum Rules with Three-Loop Corrections}",
    eprint = "hep-ph/0207183",
    archivePrefix = "arXiv",
    journal = "Eur. Phys. J. C",
    volume = "27",
    pages = "229--241",
    year = "2003",
    doi = "10.1140/epjc/s2002-01099-8"
}

@article{Drukarev:1990,
    author  = {Drukarev, E. G. and Levin, E. M.},
    title   = {The QCD sum rules and nuclear matter},
    journal = {Nuclear Physics A},
    volume  = {511},
    number  = {3-4},
    pages   = {679--700},
    year    = {1990},
    doi     = {10.1016/0375-9474(90)90117-5}
}

@article{Gubler:2019,
    author  = {Gubler, Philipp and Satow, Daisuke},
    title   = {Recent Progress in QCD Condensate Evaluations and Sum Rules},
    journal = {Progress in Particle and Nuclear Physics},
    volume  = {106},
    pages   = {1--67},
    year    = {2019},
    doi     = {10.1016/j.ppnp.2019.02.005},
    eprint  = {1812.00385},
    archivePrefix = {arXiv},
    primaryClass  = {hep-ph}
}

@article{AokiEtAl2012,
    author = {Aoki, Sinya and Doi, Takumi and Hatsuda, Tetsuo and
              Ikeda, Yoichi and Inoue, Takashi and Ishii, Noriyoshi and
              Murano, Keiko and Nemura, Hidekatsu and Sasaki, Kenji},
    title = {Lattice {QCD} approach to Nuclear Physics},
    journal = {Prog. Theor. Exp. Phys.},
    volume = {2012},
    pages = {01A105},
    year = {2012},
    doi = {10.1093/ptep/pts010},
    eprint = {1206.5088},
    archivePrefix = {arXiv},
    primaryClass = {hep-lat}
}

\end{document}